\documentclass[runningheads]{llncs}
\usepackage{graphicx}
\usepackage{svg}
\usepackage{enumitem}
\usepackage{url}
\usepackage{hyperref}
\usepackage{float}
\usepackage{multirow}
\begin{document}
\title{LooperGP: A Loopable Sequence Model for Live Coding Performance using GuitarPro Tablature}
\titlerunning{LooperGP: A Loopable Sequence Model for Live Coding}
%

\author{Sara Adkins\inst{1} \and
Pedro Sarmento\inst{1}\orcidID{0000-0002-4518-0194} \and
Mathieu Barthet\inst{1}\orcidID{0000-0002-9869-1668}}
\authorrunning{S. Adkins et al.}
%
 \institute{Centre for Digital Music, Queen Mary University of London, United Kingdom 
 \email{sara.adkins65@gmail.com, \{p.p.sarmento, m.barthet\}@qmul.ac.uk}\\
}
\maketitle              
\begin{abstract}
Despite their impressive offline results, deep learning models for symbolic music generation are not widely used in live performances due to a deficit of musically meaningful control parameters and a lack of structured musical form in their outputs. To address these issues we introduce LooperGP, a method for steering a Transformer-XL model towards generating loopable musical phrases of a specified number of bars and time signature, enabling a tool for live coding performances. We show that by training LooperGP on a dataset of 93,681 musical loops extracted from the DadaGP dataset \cite{Sarmento2021}, we are able to steer its generative output towards generating 3x as many loopable phrases as our baseline. In a subjective listening test conducted by 31 participants, LooperGP loops achieved positive median ratings in originality, musical coherence and loop smoothness, demonstrating its potential as a performance tool. 

\end{abstract}
\section{Introduction}
Sequence models such as the Pop Music Transformer \cite{Huang2020} and Transformer-XL \cite{Dai2019} are able to generate symbolic polyphonic music that maintains a coherent musical structure over several minutes of audio. However, these types of models present several issues that prevent them from being viable in the context of a live performance. They offer no control over important musical parameters such as key signature, time signature, instrumentation, genre and duration. While the outputs generated may be musically sound, they are difficult to incorporate into a performance if these structural music parameters are uncontrollable. Furthermore, the slow speed of inference makes it very difficult to develop a usable interface for running sequence models in traditional performance settings.

Live coding, an artform in which a performer writes code that synthesizes music in real time, has the potential to be a performance environment in which sequence models are viable. The live coding language can be thought of as a score notation tool, where edits are evaluated in real time \cite{Magnusson2019}. In a typical performance, the performer will alter the score by generating a symbolic pattern through algorithmic composition techniques \cite{Brown2009} or hard-coding. The pattern is then typically added to the audio mix and repeated as a loop \cite{Wiggins2010}. Rather than traditional pattern generation techniques, a sequence model could be used to generate pattern options for the performer to choose from.

In this paper, we address the issue of controllable generative music by presenting a training and inference algorithm for steering a Transformer model towards generating loopable phrases of a specified length, key and time signature. We then evaluate the success of our loop model, coined as LooperGP, by comparing the number of viable loops generated to a baseline model. Finally, we present results from a listening test that compares participants' ratings of creativity, likability, coherence and smoothness between human- and machine-made loops in order to subjectively evaluate LooperGP and identify areas for improvement. Our loop model and inference controls are intended to enable the use of deep generative models in a live coding context by addressing the constraints of ensuring model outputs fit within the context of an ongoing musical performance. LooperGP evaluations in live performance will be explored in future work.



\section{Related Work}
\subsection{Sequence Models for Music Generation}
A variety of deep learning architectures such as Recurrent Neural Networks, Variational Autoencoders and Generative Adversarial Networks have been used to generate symbolic music \cite{Briot2020}, \cite{Ji2020}, and Transformers have been particularly successful at modeling temporal dependencies \cite{Dai2019}. As demonstrated by the Music Transformer in \cite{Huang2020}, the self-attention mechanism of the Transformer architecture allows the model to refer back to previously generated motifs and phrases, making it a promising architecture for symbolic music generation. The Pop Music Transformer paper further shows that the way in which musical data is stored can have a significant effect on output quality. For instance, \cite{Huang2020}'s results demonstrated that encoding musical knowledge such as metrical position into the training data resulted in better beat and meter salience.

\subsection{Controllable Music Generation}
A large barrier stifling the adoption of symbolic music generation systems in mainstream music creation is a lack of control. Often the only available control is the temperature parameter \cite{Ackley1985}, used to select a token from the model's probability distribution. One exception to this is the DeepBach system \cite{Hadjeres2016}, which allows the user to impose constraints on generation by fixing specific notes in a Bach-style chorale. Given these note constraints, the bi-directional long short-term memory (LSTM) model fills in the rest of the chorale using Gibbs sampling. This level of customization would be useful in a live coding context for generating phrases with a good loopable cadence, as start and endpoint notes could be hard-coded. However, the DeepBach method is constrained to a fixed receptive field and has only been tested on Bach chorales, potentially limiting its use in performance of modern genres.

Vanilla Transformers often expand upon a motif they are primed with, but this behavior is not guaranteed. The Theme Transformer in \cite{Shih2022} introduces a new model architecture that conditions the Transformer encoder on a specific motif/theme. The training data is manipulated such that ``theme-composition pairs" are extracted from each training example by clustering similar measures together. By training the model on these pairs, it learns how motifs are expanded upon throughout a piece and is rewarded for repeating and varying themes. This ability to expand upon existing material is useful in live performance, as it ensures generated output is related to existing music material.

\vspace{-0.15cm}
\subsection{Live Coding}
\cite{Nilson2007} defines live coding as ``the act of programming a computer under concert conditions," where we are specifically interested in programs that produce musical output. While traditional music performance involves specifying a stream of music on a note by note  or chord by chord basis, live coding involves performing at a higher level of abstraction through score-level control over the creation of music \cite{Nilson2007}. A large focus of live coding practice is on control over generative processes; a performer controls the parameters of a generative algorithm that creates the musical output such that writing software becomes an act of performing\cite{Brown2009}. 

Prior work which focused on incorporating deep learning (DL) into live coding performances includes Cibo \cite{Stewart2020}, a text-based sequence-to-sequence model that generates Tidal Cycles \cite{Wiggins2010} code, and RaveForce \cite{Lan2019}, a pipeline for translating synthesizer parameters derived from a DL model to a SuperCollider server using Open Sound Control (OSC) \cite{McCartney1996}.

\vspace{-0.15cm}
\subsection{DadaGP Dataset}
A generative model used in a live coding context will ideally be capable of outputting music from a variety of modern genres to match the styles typically found at these types of performances. The DadaGP dataset \cite{Sarmento2021} contains over 26,000 songs covering 700 genres and a dedicated encoder represent songs as tokens compatible with sequence models. The dataset emphasizes rock and metal, but also contains jazz, pop, classical and electronic dance music (EDM). Furthermore, expressive performance information for string instruments such as tremolo, palm muting and hammer-ons/pull-offs are included in the token set, giving it an advantage over MIDI. The song diversity and token expressivity of the DadaGP dataset make it an ideal training corpus candidate to base music generation on for live coding performances.

\cite{Sarmento2021} also presents a Transformer-XL model with 8 attention heads and 12 self-attention heads trained on the DadaGP dataset. We use this model as a baseline and starting point for our LooperGP model presented in the next section.

\section{Methodology}

The Theme Transformer in \cite{Shih2022} demonstrates promising results in reorganizing training data to steer the model towards certain types of outputs. The Pop Music Transformer in \cite{Huang2020} showed that including music information retrieval (MIR) features such as beat position, bar position and chord labels in the tokens improved higher level music features for the generated content like beat salience. Combining these ideas, we extract ``loopable phrases" from the training data to create a new training corpus consisting only of segments that loop naturally. Each loop will be stored as an ordered list of DadaGP tokens \cite{Sarmento2021}, where the tokens describe the score and performance details for each activated instrument in the loop. The goal is to further steer the model towards generating loopable phrases during inference, so they can be incorporated into a live coding performance. 

\subsection{Defining a Loop}
\begin{figure}
\centering
  \includegraphics[width=.7\linewidth]{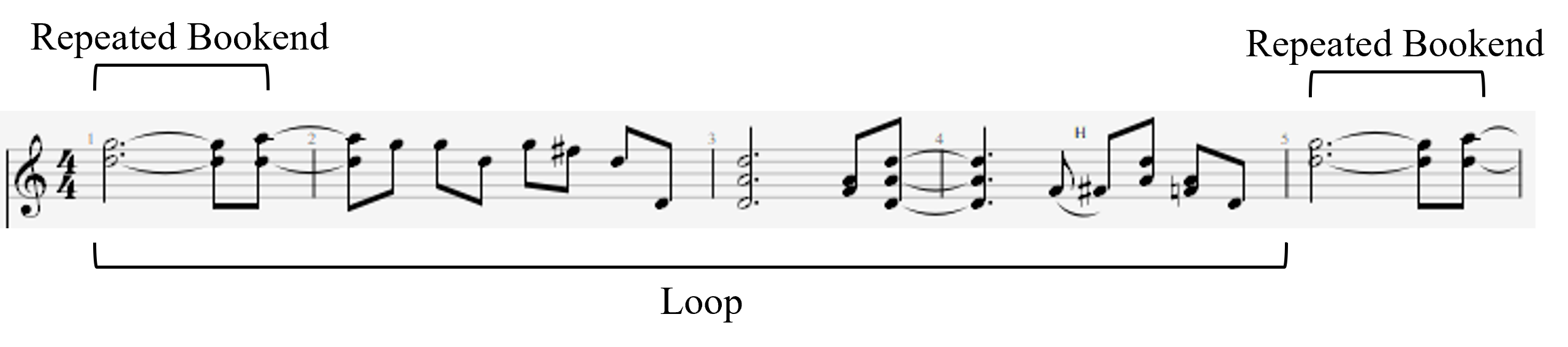}
  \caption{An example loop from AC/DC's ``You Shook Me All Night Long." Only the guitar part is shown, though the loop is multi-track in the DadaGP dataset.}
  \label{fig:loop_ex}
\end{figure}

Previous work on loop extraction has classified a loop as a short segment of audio that transitions seamlessly when repeated \cite{Ramires2020}, \cite{Chandna2021}. Seamless ``loopability" is ultimately a subjective consideration from listeners, but we can focus on the structural attributes of the composition to identify repeatable sections. For our purposes, we consider a segment of a song to be loopable if it is bookended by a repeated phrase of a minimum length\footnote{We focus here on loops where the exact same content is repeated, but it is worth noting that a more general definition could encompass loops where certain types of musical variations can occur across repetitions (e.g. modulation).}. An example loop from AC/DC's ``You Shook Me All Night Long" is shown in Figure \ref{fig:loop_ex}.

\vspace{-0.8cm}

\subsection{Extracting Loops}
\label{Loop_extract_sec}
To identify all the loops in a given song, we must first identify all the repeated phrases, or ``bookends" as shown in Figure \ref{fig:loop_ex}. To extract the repetitions from a given song, we use the correlative matrix approach presented in \cite{Hsu2001}. The DadaGP dataset \cite{Sarmento2021} is provided in a text format designed for the multi-track tablature editing software Guitar Pro. A Guitar Pro formatted song is first converted into a list $M$ of active note sets as shown in Equation \ref{eq:melody_list}. $N(t, d)$ represents the set of all active notes at tick time $t$ with duration $d$, such that $t_0 = 0$ and $t_n = t_{n-1} + d_{n-1}$.
\begin{equation}
    \label{eq:melody_list}
    M = [N(t_0, d_0),N(t_1, d_0),...,N(t_n, d_n)]
\end{equation}

Let $C_s$ and $D_s$ be $L \times L$ matrices for song $s$ with melody line $M_s$, where $L$ is the length of $M_s$. $C_s$ is populated as a correlation matrix for notes as shown in Equation \ref{eq:corr_mat}, where $C_s(i,j)$ represents the number of notes in the repeated segment that ends at both $M_s[i]$ and $M_s[j]$. Similarly, $D_s(i,j)$ is a correlation matrix  that stores the duration (in ticks) of a repeated segment that ends at both $M_s[i]$ and $M_s[j]$.
\begin{equation}
    \label{eq:corr_mat}
    C_s[i,j] = C_s[i-1,j-1] + 1 \textrm{ if } M_s[i] = M_s[j] \textrm{ else } 0
\end{equation}

The matrices $C_s$ and $D_s$ allow us to identify loop candidates and filter repeated phrases by both duration and note length. The recursive nature of the correlation matrix ensures that for any given cell $C_s[i,j] > 0$; it follows that $C_s[i-1,j-1] = C_s[i,j] - 1$. Therefore, we can identify the starting points of all repetitions of a minimum length $L_{min}$ by searching for cells in the correlation matrix where $C_s[i,j]=L_{min}$. The two starting indices of the repeated phrase $[s_i:s_j]$ are calculated in Equation \ref{eq:corr_start}, and form a potential loop $p_{i,j}$ as defined in Equation \ref{eq:phrase_def}.
\begin{equation}
    \label{eq:corr_start}
    [s_i,s_j] = (i - L_{min}, j - L_{min}) \textrm{ iff } C_s[i,j]=L_{min}
\end{equation}
\begin{equation}
    p_{i,j} = M[s_i:s_j]
    \label{eq:phrase_def}
\end{equation}

Given a list of all potential loops $p_{i,j}$ and their corresponding endpoints $[s_i, s_j]$, we can filter out loops that are trivial, too lengthy, or have repeated bookends that are too short. The following length parameters are configurable: 
\begin{itemize}
    \item Minimum Repetition Notes ($L_{min}$): minimum number of notes in bookend
    \item Minimum Repetition Beats ($RD_{min}$): minimum number of beats in bookend
    \item Minimum Loop Bars ($LB_{min}$): minimum number of bars between bookends
    \item Maximum Loop Bars ($LB_{max}$): maximum number of bars between bookends
\end{itemize}

We filter for $LB_{min}$ and $LB_{max}$ by comparing the timestamp difference between $M[s_i]$ and $M[s_j]$, and converting this tick duration to number of bars using the time signature metadata for the song. Next, we filter for $RD_{min}$ by locating the end of each loop bookend. Given a loop with endpoints $[s_i, s_j]$, we can find the end of the repeated sub-phrase by traversing along the diagonal of $C_s$ until we hit a decrease in cell value as shown in Equation \ref{eq:rep_dur1}. This decrease indicates that the matching between $s_i$ and $s_j$ has ended, and we have found the end of the bookend. We can then use Equation \ref{eq:rep_dur2} to lookup the duration of the bookend in ticks and filter out loops with duration shorter than desired.
\begin{equation}
    \label{eq:rep_dur1}
    n_{end} = n | (C_s[i+n,j+n] < C_s[i+n-1,j+n-1])
\end{equation}
\begin{equation}
    \label{eq:rep_dur2}
    D_{ticks} = D_s[i+n_{end}-1,j+n_{end}-1]
\end{equation}

A final filtering strategy to improve extracted loops' musical interest is to filter by note density, in order to filter out loops made up primarily of rests or long held notes. We defined note density ($\rho$) for a loop $p_{i,j}$ as the average number of note onsets in a measure, scaled by the number of instrumental tracks. This definition is formalized in Equation \ref{eq:density}, where $B$ is the number of bars in the loop and $T$ is the number of instrument tracks. A $\rho$ value of 4.0 for instance would mean each track in $p_{i,j}$ has on average four note onsets per measure.
\begin{equation}
    \label{eq:density}
    \rho_{p_{i,j}} = len(p_{i,j}) \times \frac{T}{B}
\end{equation}

In addition to loops extracted using the correlative matrix algorithm, we can also identify built-in loops by simply searching for repeat signs in the training corpus and filtering by the duration in beats. The full loop dataset is then made up of both extracted and built-in loops. Table \ref{tab_param_data} shows how different combinations of loop extraction parameters affect the number of loops identified. We chose the bolded parameter settings (total of 93,681 loops), as it was the median of all the parameter combinations explored.

\renewcommand{\arraystretch}{0.8}
\begin{table}[htbp]
\caption{Loop Extraction Parameter Statistics.}
\begin{center}
\begin{tabular}{|c|c|c|c|c|c|}
\hline
\textbf{Bars} & \textbf{Density} & \textbf{L min} & \textbf{RD min} & \textbf{Loops} &  \textbf{Avg. Loops Per Song}\\
\hline
4-8 & 4 & 4 & 4 & 79,066 & 3.02 \\
\hline
4-8 & 3 & 4 & 4 & 80,557 & 3.008 \\
\hline
4-8 & 4 & 2 & 4 & 82,244 & 3.14 \\
\hline
4-8 & 3 & 2 & 4 & 83,787 & 3.20 \\
\hline
4-16 & 4 & 4 & 4 & 88,842 & 3.34 \\
\hline
4-16 & 3 & 4 & 4 & 90,512 & 3.46 \\
\hline
4-8 & 4 & 4 & 2 & 92,325 & 3.52 \\
\hline
4-16 & 4 & 2 & 4 & 92,422 & 3.53 \\
\hline
\textbf{4-8} & \textbf{3} & \textbf{4} & \textbf{2} & \textbf{93,681} & \textbf{3.59} \\
\hline
4-16 & 3 & 2 & 4 & 94,138 & 3.60 \\
\hline
4-8 & 4 & 2 & 2 & 99,331 & 3.79 \\
\hline
4-8 & 3 & 2 & 2 & 100,930 & 3.89 \\
\hline
4-16 & 4 & 4 & 2 & 104,910 & 4.01 \\
\hline
4-16 & 3 & 4 & 2 & 106,607 & 4.08 \\
\hline
4-16 & 4 & 2 & 2 & 113,547 & 4.34 \\
\hline
4-16 & 3 & 2 & 2 & 115,347 & 4.41 \\
\hline
\end{tabular}
\label{tab_param_data}
\end{center}
\end{table}

\vspace{-1.3cm}

\begin{table}[htbp]
\caption{Loop Dataset Statistics.}
\begin{center}
\begin{tabular}{|c|c|c|c|c|}
\hline
\textbf{Data Format} & \textbf{Files} & \textbf{Max Tokens Per Song} & \textbf{Avg. Tokens Per Song } & \textbf{Size} \\
\hline
Hard & 66,637 & 12,253 & 1,376 & 9.13GB \\
\hline
Barred & 23,024 & 43,954 & 2,270 & 5.19GB \\
\hline
DadaGP & 26,158 & 51,997 & 4,456 & 9.56GB \\
\hline
\end{tabular}
\label{tab_loop_data}
\end{center}
\end{table}

\vspace{-1.2cm}

\subsection{Training}
\label{training_section}
We experimented with two different loop training dataset formats. The first format (“Barred Repeats”) surrounds each loop in a given song with \verb|repeat_open| and \verb|repeat_close| tokens, then concatenates these barred loops into a single file. The second format (“Hard Repeats”) stores each loop in a separate file, where the loop is manually repeated 4 times. We hypothesized that by training on the Barred Repeats dataset format, the model would learn where to place repeat open/close tokens during generation to best form musically coherent loops. The Hard Repeats model may better learn smooth transitions due to the multiple repetitions of each loop. However, as shown in Table \ref{tab_loop_data}, the Hard Repeats storage format is much less space efficient than Barred Repeats, and therefore takes longer to run through an epoch during training. We used the Pop Music Transformer-XL architecture presented in \cite{Sarmento2021} to train LooperGP. For each data format, we trained one model from scratch and a second starting from epoch 200 of the DadaGP model. The pretrained DadaGP model has already learned how to create structurally coherent music, but early testing revealed that it tended to switch to different thematic sections at random. By training this model further on a loop dataset, we hoped to teach the model to bias its generation towards coherent loops that we can then extract using the filtering procedure presented in Section \ref{Loop_extract_sec}. The results of the different training configurations are covered in the results section. For all four models, we used a learning rate of 0.0001, with 15\% dropout and an Adam optimizer as used by \cite{Sarmento2021}.

\vspace{-0.3cm}

\subsection{Controllable Inference}
\vspace{-0.1cm}
In a live coding performance, it is important that the performer has control over the musical parameters of generation. In order to fit a generated loop into an ongoing performance, the loop must match the key and time signature of what is currently playing. It is also essential to control the length of the loop in bars, to ensure the phrase lines up (or intentionally doesn't line up) with other patterns that are currently active in the set. 

\begin{figure}
  \centering
  \includegraphics[width=.7\linewidth]{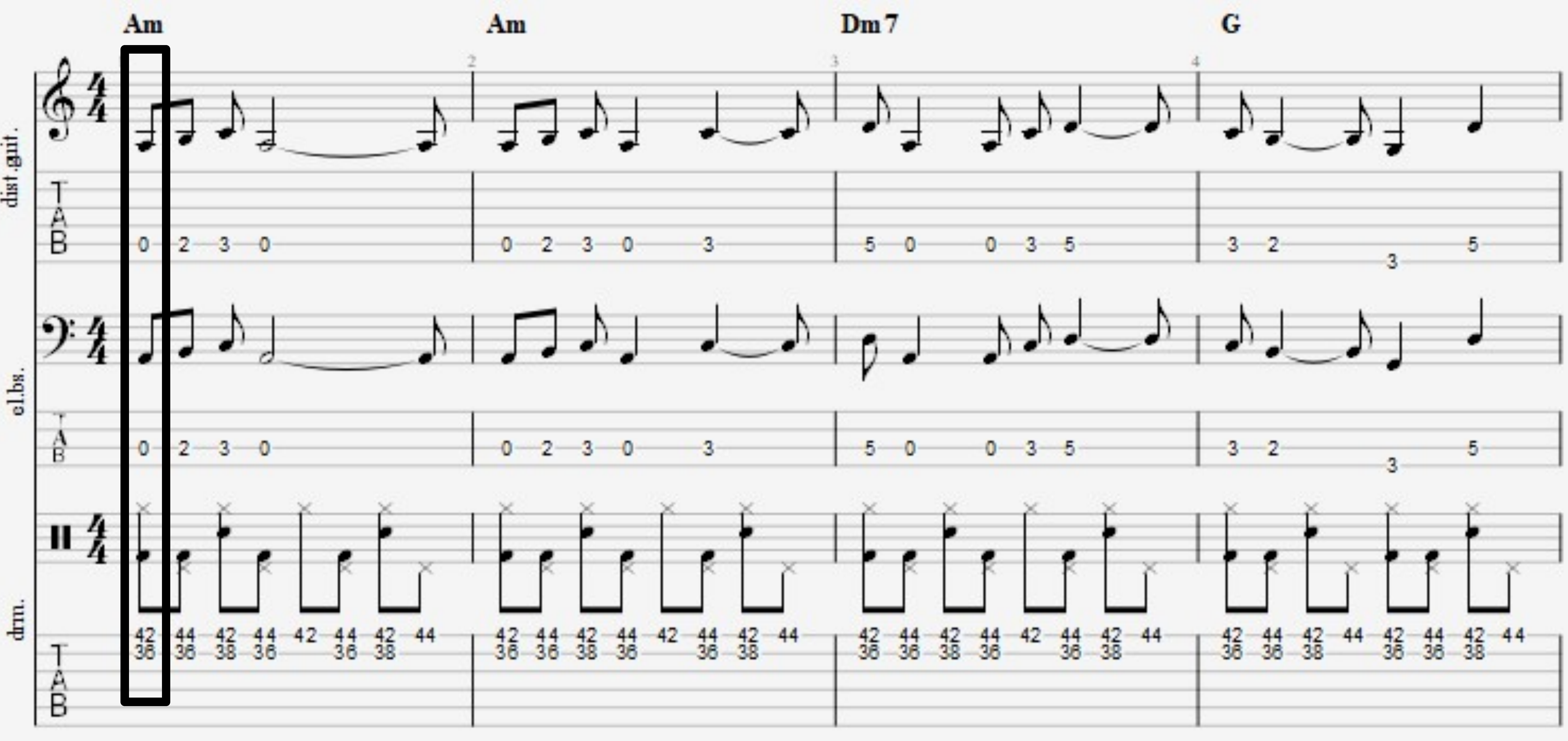}
  \caption{Example output generated from the DadaGP model, with primer shown in the black box. Chord symbols were analyzed by hand.}
  \label{fig:primer_ex}
\end{figure}

\cite{Sarmento2021} observed that priming inference with one or more notes from the desired instruments forming the root chord of the key is generally successful at generating music in that key and instrumentation, at least over a short duration. Figure \ref{fig:primer_ex} shows how priming with an eighth note of drums (unpitched), guitar (A3) and bass (A2) results in the model continuing with the established instrumentation in the key of A minor for several measures.

\vspace{-0.2cm}

\vspace{-0.4cm}

We further explored how to control time signature and duration during inference. The DadaGP Transformer model operates by generating a set number of tokens \cite{Sarmento2021}, but these tokens do not have any relationship to duration. To address this unpredictability, we modified the inference method to keep track of cumulative duration by summing up the tick values of the \verb|wait:xxx| tokens and artificially inserting \verb|new_measure| tokens after a specified amount of ticks to enforce a measure duration. If the tick duration of a wait token crosses a bar boundary, we shorten it to fit within the measure. We also keep a cumulative count of the \verb|new_measure| tokens, and artificially insert an ``end" token to replace the last \verb|new_measure| token when the desired number of bars is reached. A shortcoming of this approach is that it does not allow us to differentiate between time signatures with the same measure duration, e.g. between 3/4 and 6/8. 

After LooperGP has generated a desired number of bars, the next step is to extract loops using our filtering procedure. By setting the $LB_{min}$ and $LB_{max}$ parameters to the same value, we can specify the exact length of loops to output. As LooperGP was trained specifically on loop data, the generated outputs should contain significantly more loops compared to the baseline model. We compare the loop generation rate between different versions of LooperGP and the baseline DadaGP model in the results section.

\section{Results}
\subsection{Training Performance}
\begin{figure}
  \centering
  \includegraphics[width=0.55\linewidth]{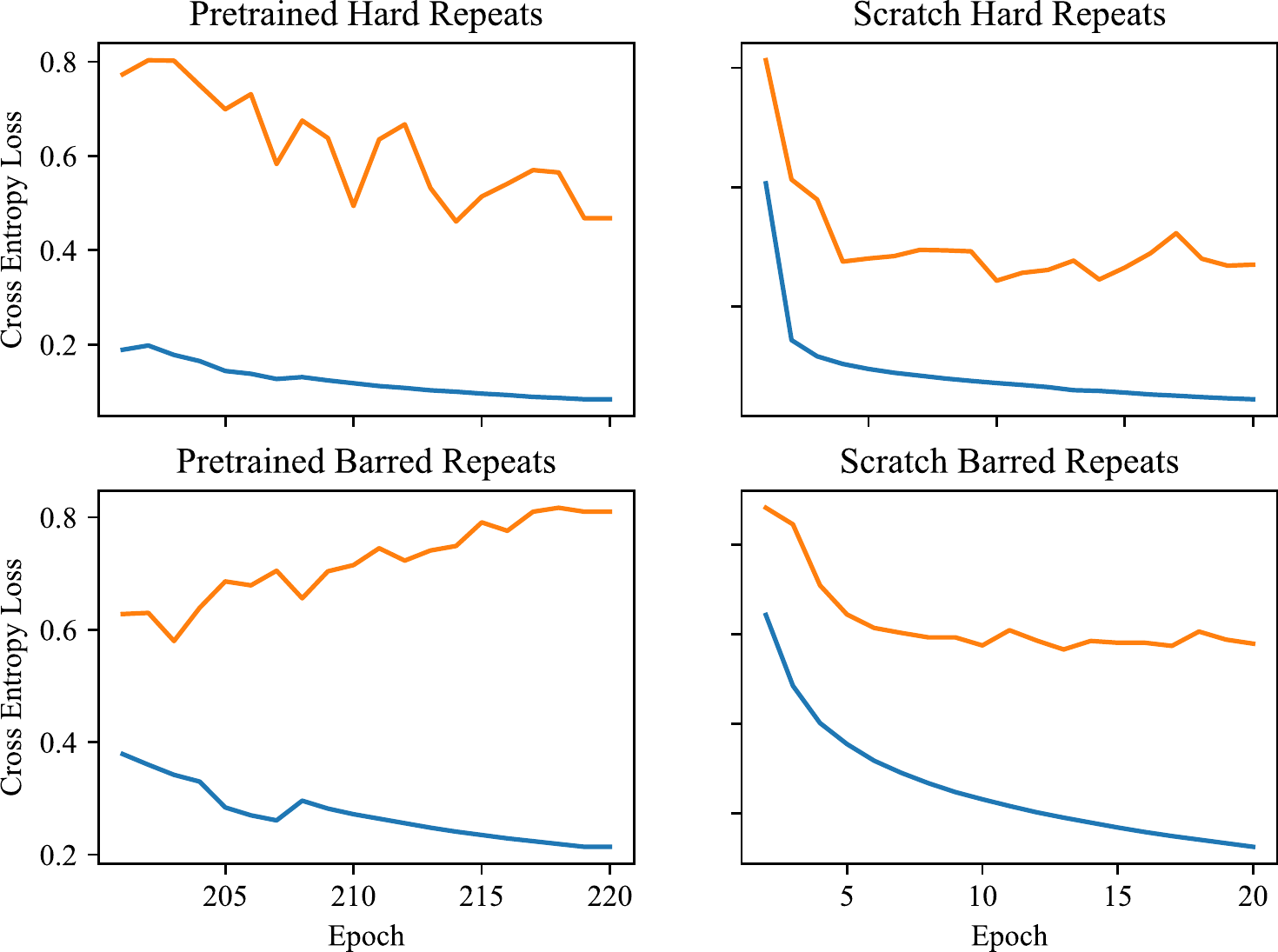}
  \caption{Training loss (blue) and Validation loss (orange) across 20 epochs for the 4 LooperGP model configurations}
  \label{fig:training}
\end{figure}

The four model configurations described in Section \ref{training_section} were each trained for 20 epochs (due to resource constraints) with an 85/15 training/validation split on the loop dataset. Cross-entropy training and validation loss curves are shown in Figure \ref{fig:training}. As expected, the pretrained model configurations have a much flatter loss curve than the scratch models, as they have already converged on the original DadaGP dataset. Validation loss decreases over time in all but the Pretrained Barred Repeats model. 

\vspace{-0.2cm}

\subsection{Loop Performance}
To evaluate the quality of output generated by each of the model configurations, we generated 25 excerpts from each model and calculated the average number of loops extracted per excerpt and the average loop note density as defined in Equation \ref{eq:density}. Each model was configured to generate 16 bars of 4/4 music, primed with a random tempo and instrument configuration extracted from the training data. We then ran the loop extraction algorithm from Section \ref{Loop_extract_sec} with the following parameters: $L_{min}=4$, $RD_{min}=2$, $LB_{min}=4$ and $LB_{max}=4$, chosen as the median of Table \ref{tab_param_data}

The results from the inference analysis are shown in Table \ref{tab_inference_stats}. It is immediately clear that the Hard Repeats model trained from scratch was not able to generate meaningful loops; very few were extracted and the average note density was extremely low. Manually viewing the generated results from this model revealed most of the outputs were comprised of long held notes and excessive rests, indicating it was not able to sufficiently learn musical phrases. The Pretrained Hard Loops model did not have this issue with excessive rests, however it failed to perform better than the baseline DadaGP model, generating only 1 more valid loop across the total 25 excerpts. 

\vspace{-0.3cm}

\begin{table}[htbp]
\caption{Loop Extraction Statistics Between Models.}
\begin{center}
\begin{tabular}{|c|c|c|c|}
\hline
\textbf{Model} & \textbf{Loops Found} & \textbf{Avg. Loops} & \textbf{Avg. Note Density} \\
\hline
DadaGP Baseline & 20 & 0.80 & 6.03 \\
\hline
Pretrained Hard & 21 & 0.84 & 6.84 \\
\hline
Scratch Hard & 3 & 0.12 & 0.25 \\
\hline
Pretrained Barred & 88 & 3.52 & 6.96 \\
\hline
Scratch Barred & 77 & 3.08 & 9.80 \\
\hline
\end{tabular}
\label{tab_inference_stats}
\end{center}
\end{table}

\vspace{-0.8cm}

The Barred Repeats models both performed well above the baseline in terms of number of loops generated. In both cases three or more 4-bar loops were extracted from each 16 bars of generation, with high average note density. 
While these results are promising, they do not give us any indication as to the musical quality of the extracted loops. To prepare stimuli for a subjective listening test, we continued to train the Pretrained Barred Repeats model, stopping at epoch 40 with a Cross-Entropy loss of 0.148.

\subsection{Subjective Evaluation}
To evaluate the subjective quality of loops generated by LooperGP, we conducted an online listening study which had participants evaluate machine-generated loops and human-composed loops extracted from the training data. Participants listened to 50 excerpts, each containing a 4-bar loop that repeated four times. 25 of the excerpts were human-composed, and the other 25 were machine-generated\footnote{Link to listening test excerpts: \url{https://drive.google.com/drive/folders/1I0MCPYjj8nXqKkmDN-d-C2ETOHJpCZyn?usp=share_link}}. Excerpts were presented to participants in a random order, without revealing the generation method. Participants were asked the following list of questions about each excerpt, and also filled out the Goldsmiths Music Sophistication Index (MSI) survey \cite{goldsmith2014}:
\begin{enumerate}
    \item Have you heard the music in this excerpt before? (Y/N)
    \item Do you think the music in this loop has been produced by a human or by a machine? (Human/Machine)
    \item Please rate to what extent you agree or disagree with the following statements: (7-point Likert scale)
    \begin{itemize}[label={}]
        \item Q1: The music in this loop is original and musically creative.
        \item Q2: I like the music in this loop.
        \item Q3: Within the loop, the music is coherent rhythmically, melodically and harmonically.
        \item Q4: The transition between loop repetitions is smooth rhythmically, melodically and harmonically.
    \end{itemize}
    \item Briefly explain the reasoning behind your ratings. You may list keywords rather than using full sentences.
\end{enumerate}

\vspace{-0.7cm}
\subsubsection{Listening Test Excerpts}
The 25 human-made loops were randomly sampled from the set of all 4-bar loops extracted from the DadaGP training data. The other 25 excerpts were machine-generated by the LooperGP model. To generate an excerpt, a primer tempo, instrumentation and 1st note were randomly sampled from the 4-bar loop DadaGP dataset, then inference was run for 16 bars of 4/4. Next, all 4-bar loops were extracted from the generated output and a random one was chosen as the final excerpt. This procedure was repeated 25 times. All excerpts were rendered as MP3s (variable 260 kbps) using the default virtual instruments in GuitarPro 7. Participants were instructed to focus on the quality of the music composition and not on the quality of the virtual instruments or the music production mix when answering questions.

\vspace{-0.4cm}
\subsubsection{Participants}
We recruited 31 participants in total, six of whom responded to a department-wide participant invitation email, and 25 from the data collection website Prolific. The inclusion criteria were access to headphones, normal hearing, no amusia (tone deafness), and having music as a hobby. Participants were roughly evenly split between genders, with 17 male and 14 female participants.

\vspace{-0.4cm}
\subsubsection{Ordinal Data Transformation}
\begin{figure}
  \begin{center}
  \includegraphics[width=0.5\linewidth]{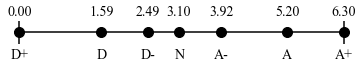}
  \caption{7-point Likert scale values converted to a continuous scale (Snell method \cite{Wu2007}).}
  \label{fig:likert_scale}
  \end{center}
\end{figure}

The Likert scale data collected from the four statements in Question 3 is ordinal data. Opinions are mixed as to whether or not Likert data can be treated as continuous data in statistical tests (\cite{Bishop2015} and \cite{Sullivan2013}). The mean value is not necessarily a useful metric, as participants may not assume a constant distance between Likert items \cite{Wu2007}. To work around this issue, we implemented the Snell scaling procedure proposed in \cite{Wu2007}, which uses Maximum Likelihood Estimation to place the ordinal Likert categories as points on a continuous scale so that mean-based parametric methods may be applied. Applying their algorithm to our listening test data, we mapped the seven Likert values (Strongly Disagree, Disagree, Slightly Disagree, Neutral, Slightly Agree, Agree, Strongly Agree) to the values shown in Figure \ref{fig:likert_scale}.


\subsubsection{Rating Results and Mixed Effects Models}
We used a linear mixed-effects model (LMM) \cite{meteyard2020} to test the effect of the generation type on the results from the survey questions. A LMM was chosen as it is suitable for repeated measures and individual differences can be modeled by using different random intercepts for each participant.
Participants' MSI scores \cite{goldsmith2014} were included as an additional independent variable to evaluate the relationship between music background and ratings. Finally, we used participant ID (PID) as a random effect to account for variation in rating tendencies between individual participants. The mixed effects model was implemented using the random intercept model from the StatsModel 0.13.5 Python library \cite{statsmodels}. In all four questions, generation type had a significant effect on the rating results, with $p < .001$. The MSI score did not have a significant effect on any of the dependent variables, with all $p \geq 0.22$. Table \ref{tab_mixed_ff} shows the breakdown of regression results by question.

\vspace{-0.4cm}

\renewcommand{\arraystretch}{1}
\vspace{-1.5cm}
\begin{table}[htbp]\caption{LMM Results for Fixed Effect (Coefficients, Standard Errors, and p-value) and Random Effect (PID Variance and Standard error)}
\begin{center}
\begin{tabular}{|c|c|c|c|c|c|c|c|c|c|c|c|c|}
\hline
\textbf{} & \multicolumn{4}{|c|}{\textbf{Coef.}} & \multicolumn{4}{|c|}{\textbf{Std. Err.}} & \multicolumn{4}{|c|}{\textbf{$P>|z|$}}\\ 
\hline
\textbf{} & \textbf{Q1} & \textbf{Q2} & \textbf{Q3} & \textbf{Q4} & \textbf{Q1} & \textbf{Q2} & \textbf{Q3} & \textbf{Q4} & \textbf{Q1} & \textbf{Q2} & \textbf{Q3} & \textbf{Q4}\\
\hline
\textbf{Intercept} & 3.388 & 3.463 & 4.019 & 3.966 & 0.584 & 0.581 & 0.495 & 0.643 & 0.001 & 0.001 & 0.001 & 0.001 \\
\hline
\textbf{Gen. Type} & -0.072 & -0.897 & -0.621 & -0.565 & 0.072 & 0.082 & 0.070 & 0.069 & 0.001 & 0.001 & 0.001 & 0.001  \\
\hline
\textbf{MSI} & 0.009 & 0.005 & 0.006 & 0.008 & 0.007 & 0.007 & 0.006 & 0.008 & 0.220 & 0.446 & 0.315 & 0.310 \\
\hline
\textbf{PID Var.} & 0.420 & 0.403 & 0.292 & 0.522 & 0.086 & 0.075 & 0.063 & 0.110  & {-} & {-} & {-} & {-}\\
\hline
\end{tabular}
\label{tab_mixed_ff}
\end{center}
\end{table}





\begin{figure*}
  \begin{center}
  \includegraphics[width=\textwidth]{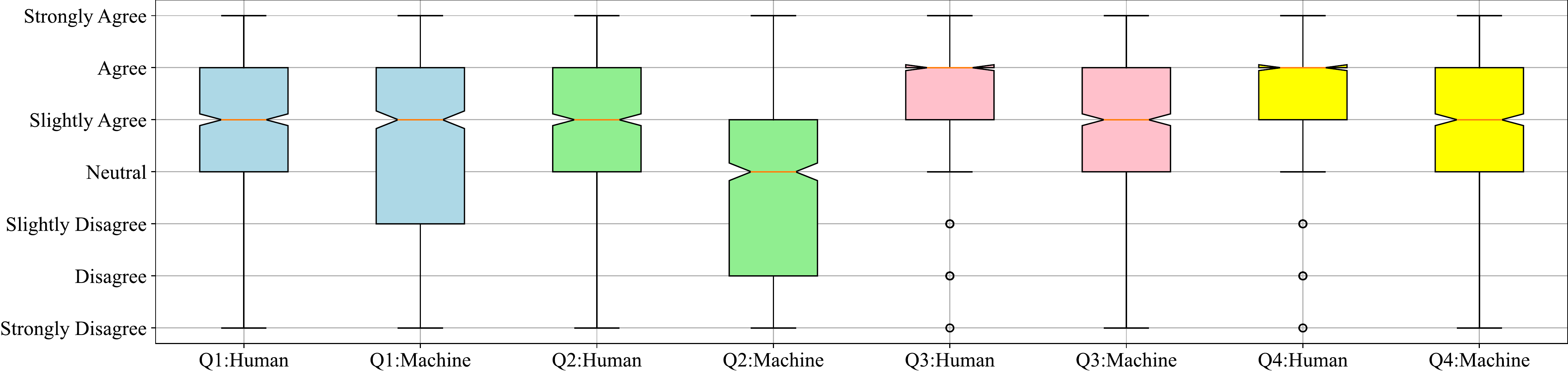}
  \caption{Comparing 7-point Likert rating distributions between human and machine generated loops for each evaluation question.}
  \label{fig:boxplot}
  \end{center}
\end{figure*}

\vspace{-0.9cm}
To visualize the rating difference between generation type, we plotted the Likert distributions shown in Figure \ref{fig:boxplot} using the original Likert data. With the exception of Q2:Machine (``I like the music in this loop"), the median of all response categories are positive. For Likert questions 2-4, the human-composed excerpts are rated approximately one category higher than the machine-generated excerpts. The median ratings for Q1 (``The music in this loop is original and musically creative") were equivalent across generation type. Mean, variance and standard deviation for the Snell-scaled \cite{Wu2007} ratings are shown in \ref{tab_likert_stats}.


\vspace{-0.3cm}

\subsubsection{Musical Turing Test Results}
We evaluated participants ability to discern between human- and machine-made loops using a logistic regression model. As with the Likert rating questions, both generation type and MSI score were used as independent variables in the model. To ensure participants' answers would not be biased towards human-made loops due to familiarity, we discarded 83 (out of 1550) datapoints where the participants indicated they recognized the music in the loop. We used the logistic regression implementation from the StatsModel Python library \cite{statsmodels} to calculate the model results. Generation type was found to have a significant effect on perceived generation source (Human or Machine), with $p < 0.01$. MSI score did not have a significant effect on perceived generation source, with $p \geq 0.28$. To further discern how well participants were able to differentiate between human- and machine-made loops, we plotted the confusion matrix in Table \ref{fig:turing_confusion}. Though generation type was found to have a significant effect on perceived generation source, the hit rate is only 10\% above chance. Participants correctly predicted generation type about 60\% of the time, and the distribution is similar for both human and machine sources.

\vspace{-0.3cm}

\vspace{-1.2cm}
\begin{table}[htbp]
\caption{Likert Rating Statistics. H: Human; M: Machine.}
\begin{center}
\begin{tabular}{|c||c|c|c|c|c|c|c|c|}
\hline
\textbf{} & \multicolumn{2}{|c|}{\textbf{Median}} & \multicolumn{2}{|c|}{\textbf{Mean}} & \multicolumn{2}{|c|}{\textbf{Var}} & \multicolumn{2}{|c|}{\textbf{Std Dev}}\\ 
\hline
\textbf{} & \textbf{H} & \textbf{M} & \textbf{H} & \textbf{M} & \textbf{H} & \textbf{M} & \textbf{H} & \textbf{M}\\
\hline
\textbf{Q1} & Slightly Agree & Slightly Agree & 4.09 & 3.38 & 2.24 & 2.62 & 1.50 & 1.62\\
\hline
\textbf{Q2} & Slightly Agree & Neutral & 3.89 & 3.00 & 2.93 & 3.04 & 1.71 & 1.74 \\
\hline
\textbf{Q3} & Agree & Slightly Agree & 4.50 & 3.88 & 1.83 & 2.55 & 1.35 & 1.60 \\
\hline
\textbf{Q4} & Agree & Slightly Agree & 4.60 & 4.03 & 1.94 & 2.73 & 1.39 & 1.65\\
\hline
\end{tabular}
\label{tab_likert_stats}
\end{center}
\end{table}

\vspace{-1.2cm}
\begin{table}[htbp]
\caption{Confusion Matrix of Ground Truth and Predicted Generation Type.}
\begin{center}
\begin{tabular}{|c|c|c|}
\hline
\textbf{} & \textbf{Human} & \textbf{Machine} \\
\hline
\textbf{Human} & 0.6 & 0.4 \\
\hline
\textbf{Machine} & 0.42 & 0.58 \\
\hline
\end{tabular}
\label{fig:turing_confusion}
\end{center}
\end{table}

\vspace{-1.3cm}

\subsubsection{Free Response Results} 
\begin{figure}
  \begin{center}
  \includegraphics[width=0.45\linewidth]{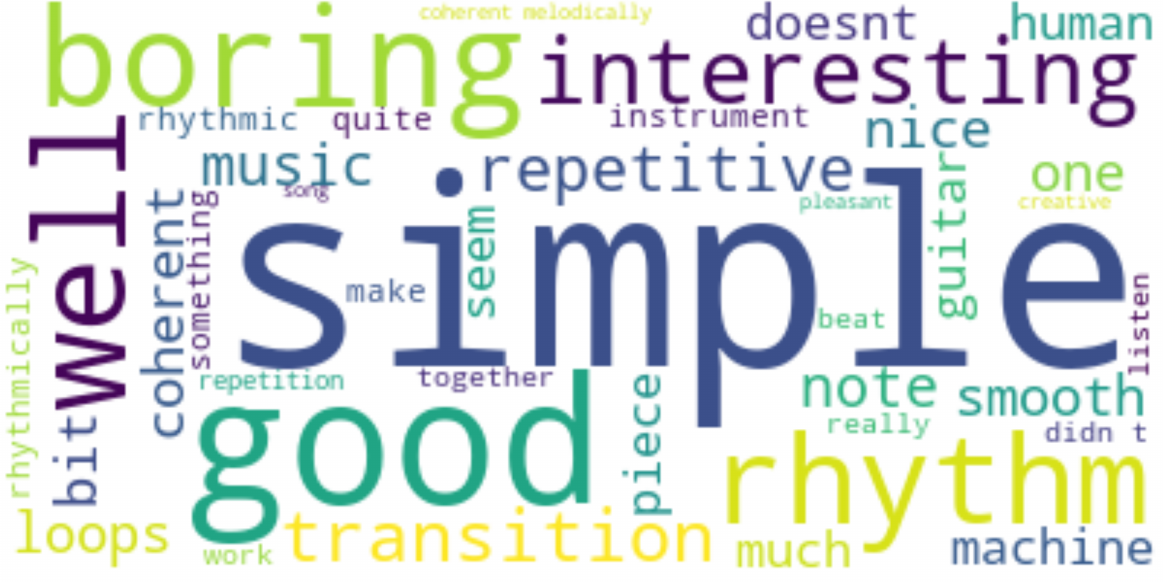} \hspace{0.25cm}
  \includegraphics[width=0.45\linewidth]{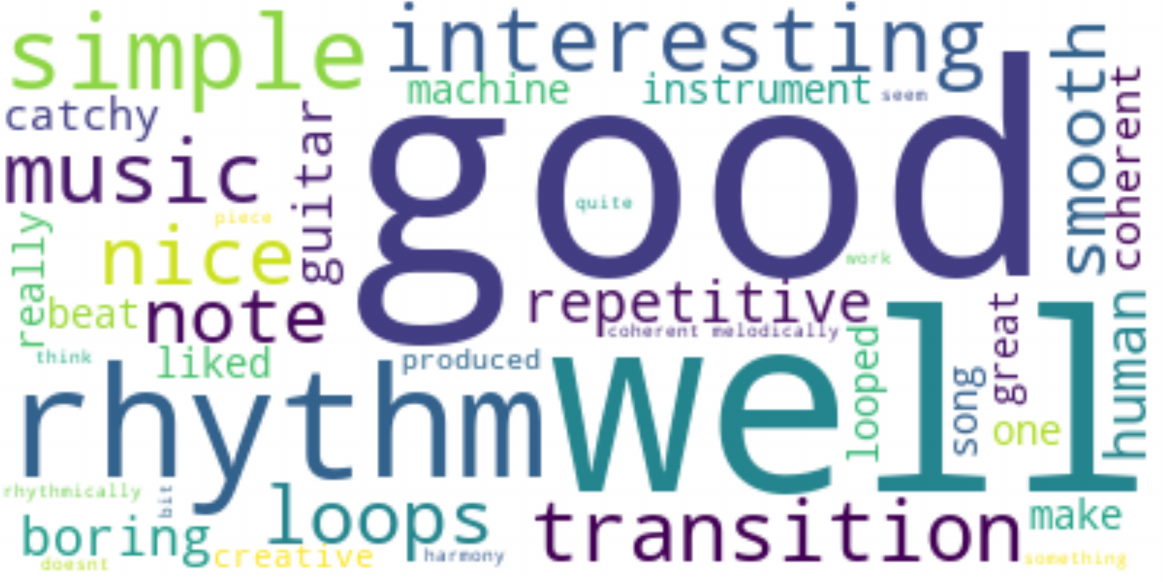}
  \caption{Frequency-based word clouds for machine-generated (left) and human-generated (right) excerpts from listening test free response.}
  \label{fig:word_cloud}
  \end{center}
\end{figure}

Participants were given a text box to explain their ratings for each excerpt in the listening test. We visualized the top 40 most frequently occurring words in answers for each generation type using a word cloud generated by the WordCloud Python library \cite{wordcloud}. The following words were used as stop words and removed from the cloud as they occurred frequently but were not informative: loop, melody, sounds and feels. As shown in Figure \ref{fig:word_cloud}, ``simple", ``boring" and ``repetitive" are used more often to describe machine excerpts than human ones. On the other hand, ``interesting", ``smooth" and ``coherent" are roughly equal size in both types of excepts. ``Good" is a top word in both categories, but is more often used to describe human excerpts.

\vspace{-1.7cm}

\section{Discussion}
Our approach to restructuring the DadaGP dataset was overall successful in encouraging loopable outputs. However, there was a surprising difference in output quality between the Hard Repeats and Barred Repeats dataset formats. The Pretrained Hard Repeats model produced the same amount of loops on average as the baseline model, indicating the model was unable to learn what features constitute a loopable phrase. The Hard Repeats model performed even worse when trained from scratch, falling into a local minimum in which the best it could do was generate rests and the occasional long note. Seemingly, the model learned to focus on the easier low density tracks while ignoring the pertinent features in the high density tracks of a song. 

The Barred Repeats model was much more successful at generating loops during inference, with both the pretrained and scratch models generating, on average, 3x as many loops as the baseline. Despite high performance on this loop metric, the validation loss of the pretrained Barred Repeats model strictly increased over time, indicating that it was over fitting the training data. Evidently the model learned to place repeat bars, but it did not learn to place them in meaningful places. Unlike the pretrained version, the scratch Barred Repeats model did converge and generated loops with a higher average note density, indicating that it was successful at learning musically meaningful loops. 


Our subjective listening test allowed us to evaluate the quality of generated loops beyond the objective note density metric. Our statistical evaluation showed a significant effect of generation type on all four Likert rating questions, indicating that the generated outputs are distinguishable from human-composed ones. However, it is promising that median ratings were positive across all questions and suggests our method has potential as a performance tool.

Breaking down the Likert ratings by question, it is peculiar that Q1, which asked for ratings of originality and musicality, had the most similar ratings across generation types. Musicality and originality are particularly human characteristics that AI models traditionally have difficulty emulating, so this result is surprising. It is possible that unexpected musical outputs were interpreted as ``original" by participants, inflating ratings for machine-made excerpts. Q2, which asks participants if they like an excerpt, had the lowest median rating for both machine and human excerpts. Though we asked participants to focus on the musical score rather than the quality of the digital instruments, for preference questions such as this one it may have been difficult for participants not to be influenced by the synthetic timbres and lack of human expression in the renditions of both the human and machine excerpts, biasing the scores negatively. Q3 and Q4, which asked about musical coherence and loop-point smoothness respectively, had very similar rating distributions as shown in Figure \ref{fig:boxplot}, hinting that perhaps participants had trouble differentiating between the meaning of musical coherence and loop-point smoothness. This may be in part caused by the presence of sub-loops within some of the excerpts. For instance, with our extraction algorithm, it is possible a 4-bar loop could actually be made up of a repeated 2-bar loop. This may have made it difficult for participants to focus on the exact loop point when answering Q4 since the amount of repetitions varied.

The musical Turing test data showed a significant effect of generation type on perceived generation srouce, but results were still close to chance for both types. While the usefulness of Turing tests in evaluating the musicality of algorithmic composition systems is disputed \cite{Ariza2009}, in our case it is a useful metric of how well LooperGP is able to mimic the loopability of the training data. In other words, it offers insight into whether LooperGP has learned to place repeat bars in convincing places. Given our results, LooperGP is able to somewhat convincingly emulate human composers, but there is room for improvement. 

The sample size for the listening test remains fairly small (31 participants who each assessed 25 items per condition, so a total of about 775 observations per condition). However, for fixed effects, power in LMMs does not necessarily increase as the total sample of observations increases \cite{meteyard2020}.

\vspace{-0.3cm}
\section{Conclusion and Future Work}
As the motivation behind LooperGP is for it be used in live coding, we shall discuss its potential in this context. Reference \cite{Brown2009} argues that for a generative process to be useful in a live coding environment, it must be succinct to call, applicable to a variety of musical circumstances, modifiable and require limited temporal scope. Given that LooperGP focuses on loopability of output and can be primed with a specific instrumentation, key/time signature and bar length, it can be configured to fit a wide variety of musical circumstances. Our algorithm requires no temporal scope, as the model does not have any temporal dependencies. A downside of LooperGP, and many other DL methods for music generation, is a lack of modification control. Once a loop has been generated, there is no way to algorithmically adjust it. However, it could be feasible to generate variations on a generated loop by passing it back through the system as a primer, as observed in \cite{Shih2022}. Finally, an inference of LooperGP can be easily generated via a succinct Python function call, as the only required parameters are key, duration and bar length.

Overall, LooperGP was able to increase the frequency of high-density loopable outputs compared to the baseline. Subjective ratings showed a positive response to the loops overall, but also identified some key areas for improvement in reducing simplicity and repetitiveness. LooperGP exhibits many of the characteristics required for a generative process to be useful in live coding, with a specific focus on generating controllable output able to fit most musical circumstances. The duration control and loopable generation of LooperGP make it a viable option for use in a live performance context, although this should be the object of further assessment with live coders.

To address the listening test comments that machine-made loops were ``repetitive" and ``boring", future work could focus on increasing the average complexity of extracted loops by adjusting the loop extraction algorithm to identify and filter out sub-loops. On the training side, more of the hyperparameter space could be explored, such as increasing dropout or decreasing layer size, to prevent the early overfitting demonstrated by some of the training setups.

Finally, an exciting area for future work is integrating the LooperGP system into a live coding framework such as Tidal Cycles \cite{Wiggins2010} or SuperCollider \cite{McCartney1996}. The OSC pipeline for communicating between Python and SuperCollider presented in \cite{Lan2019} could be used to prototype this concept without requiring the model to be exported to a new environment. It is also important that LooperGP be wrapped into a usable interface with a succinct API. Ideally, the performer should be able to cherry-pick a loop from a set of potential options generated by the model and automatically transpose it to their desired key signature. Once LooperGP has been wrapped into a performer-friendly interface, its potential as a live coding feature can be truly evaluated by incorporating it into a concert and collecting feedback from audience members and performers.

\section{Acknowledgements}
This work has been partly supported by the EPSRC UKRI Centre for Doctoral Training in Artificial Intelligence and Music (Grant no. EP/S022694/1).

%
%
%
\bibliographystyle{splncs04}
\bibliography{biblio}
\end{document}